\pdfoutput=1

\documentclass[11pt]{article}

\usepackage{acl}

\usepackage{times}
\usepackage{latexsym}
\usepackage{amsmath}
\usepackage{multirow}
\usepackage{bbding}
\usepackage{indentfirst}
\usepackage[T1]{fontenc}

\usepackage[utf8]{inputenc}

\usepackage{microtype}

\usepackage{graphicx}
\usepackage{booktabs}
\usepackage{tabularx}
\usepackage[inline]{enumitem}

\renewcommand{\thefootnote}{\fnsymbol{footnote}}

\newcommand\cdts{context-dependent text-to-SQL} 
\newcommand\cits{context-independent text-to-SQL} 
\newcommand\picard{P{\small{ICARD}}}
\newcommand\spider{Spider}
\newcommand\wikisql{WikiSQL}
\newcommand\cosql{CoSQL}
\newcommand\sparc{SparC}
\newcommand\editsql{EditSQL}
\newcommand\istsql{IST-SQL}
\newcommand\igsql{IGSQL}
\newcommand\rtsql{R$^2$SQL}
\newcommand\score{SCoRe}
\newcommand\grappa{GraPPa}
\newcommand{\ttd}{\text{\textdagger}}

\usepackage[misc]{ifsym}

\title{HIE-SQL: History Information Enhanced Network for Context-Dependent Text-to-SQL Semantic Parsing}

\usepackage{footmisc}
\author{Yanzhao Zheng$^*$ \quad Haibin Wang$^*$ \quad Baohua Dong \quad Xingjun Wang \quad Changshan Li$^\ttd$ \\
  Alibaba Group, Hangzhou, China \\
  \small{\texttt{\{zhengyanzhao.zyz,binke.whb,baohua.dbh,xingjun.wxj\}@alibaba-inc.com}}, \\ \small{\texttt{lics16@mails.tsinghua.edu.cn}}}

\begin{document}

\maketitle

\begin{abstract}
Recently, \cdts{} semantic parsing which translates natural language into SQL in an interaction process has attracted a lot of attention. Previous works leverage context-dependence information either from interaction history utterances or the previous predicted SQL queries but fail in taking advantage of both since of the mismatch between natural language and logic-form SQL. In this work, we propose a \textbf{H}istory \textbf{I}nformation \textbf{E}nhanced text-to-\textbf{SQL} model (\textbf{HIE-SQL}) to exploit context-dependence information from both history utterances and the last predicted SQL query. In view of the mismatch, we treat natural language and SQL as two modalities and propose a bimodal pre-trained model to bridge the gap between them. Besides, we design a schema-linking graph to enhance connections from utterances and the SQL query to the database schema. We show our history information enhanced methods improve the performance of HIE-SQL by a significant margin, which achieves new state-of-the-art results on the two \cdts{} benchmarks, the \sparc{} and \cosql{} datasets, at the writing time. 

\end{abstract}

\let\thefootnote\relax\footnotetext{$^*$ Equal contribution.}
\footnotetext{$^\ttd$ Corresponding author.}

\section{Introduction}

\begin{figure}[h]
    \centering
    \includegraphics[width=0.46\textwidth]{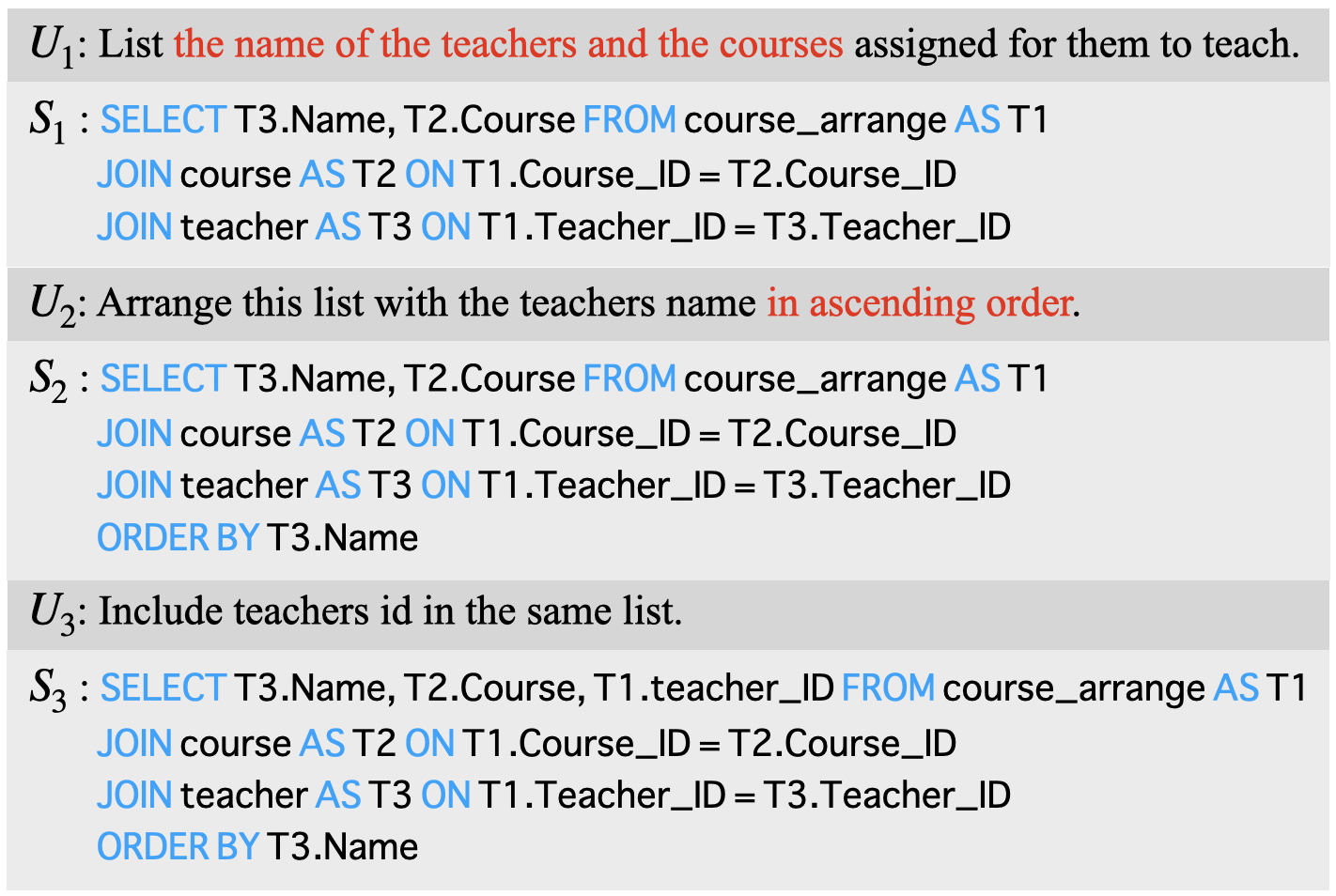}
    \caption{An example of \cdts{} interaction in \cosql{} where $U_i$ is the utterance of turn $i$ and $S_i$ is the corresponding SQL query for $U_i$. The tokens with red color are the history information that should be considered in later predictions. It is context-independent if we just consider the prediction of $S_1$.}
    \label{fg:popg1}
\end{figure}

Conversation user interfaces to databases have launched a new research hotspot in Text-to-SQL semantic parsing~\citep{DBLP:conf/emnlp/ZhangYESXLSXSR19, DBLP:conf/acl/GuoZGXLLZ19, DBLP:conf/acl/WangSLPR20, DBLP:conf/emnlp/LinSX20, DBLP:conf/acl/Xu0YZT0CPC20, DBLP:conf/acl/CaoC0ZZ020, DBLP:conf/aaai/HuiGRLLSHSZZ21, DBLP:conf/iclr/0009ZPMA21} and benefited us in industry~\citep{dhamdhere2017analyza, weir2020dbpal}. Most previous works focus on the \cits{} task and propose many competitive models. Some models~\citep{DBLP:conf/acl/WangSLPR20, scholak2021picard} even surprisingly work well on the \cdts{} task by just appending the interaction history utterances to the input. Especially, \picard{}~\citep{scholak2021picard} achieves state-of-the-art performances both in \spider~\citep{DBLP:conf/emnlp/YuZYYWLMLYRZR18}, a cross-domain \cits{} benchmark, and \cosql~\citep{DBLP:conf/emnlp/YuZELXPLTSLJYSC19}, a cross-domain \cdts{} benchmark, before our work. However, every coin has two sides. That implies underachievement of the exploration of context information in \cdts{} semantic parsing.

Compared with \cits{} semantic parsing, \cdts{} semantic parsing are more challenging since of the various types of dependence in utterances which make models vulnerable to parsing errors. As \rtsql{}~\citep{DBLP:conf/aaai/HuiGRLLSHSZZ21} considers, different context dependencies between two adjacent utterances require the model to establish dynamic connections between utterances and database schema carefully. However, context information is not only from the last utterance. Long-range dependence is also the case in \cosql{} as the prediction of $S_3$ depends on "the name of the teachers and the courses" in $U_1$ in Figure~\ref{fg:popg1}. A workable proposition for long-range dependence is to inherit context information from previous predicted SQL queries. But it is not a piece of cake to take advantage of previously predicted queries since of the mismatch between natural language and logic-form SQL. As \citet{DBLP:conf/ijcai/LiuCGLZZ20}
conclude, roughly encoding the last predicted SQL query and utterances takes the wooden spoon while easily concatenation of interaction history utterances and current utterance appears to be strikingly competitive in their evaluation of 13 existing context modeling methods.

In this paper, we propose a history information enhanced network to make full use of both history interactive utterances and previous predicted SQL queries. We first treat the logic-form SQL query as another modality with natural language. We present SQLBERT, a bimodal pre-trained model for SQL and natural language which is able to capture the semantic connection and bridge the gap between SQL and natural language. It produces general-purpose representations and supports our \cdts{} semantic parsing.

Besides, we propose a history information enhanced schema-linking graph to represent the relations among current utterance, interaction history utterances, the last predicted query, and corresponding database schema. Considering it is weird to shift a topic back and forth in an interaction, we assume that the long-range dependence is successive. For example, that $S_3$ depends on $U_1$ implies that $S_2$ does too in Figure~\ref{fg:popg1}. In that case, we can leverage the long-range dependence from the last predicted query. Therefore, unlike the previous schema-linking graph just with utterances and database schema~\citep{DBLP:conf/aaai/HuiGRLLSHSZZ21}, the last predicted query takes part in our graph. Besides, we distinguish current utterance and interaction history utterances in the schema-linking graph. We encode the schema-linking relations with Relative Self-Attention Mechanism~\citep{DBLP:conf/naacl/ShawUV18}.

In our experiments, the proposed methods of SQLBERT and the history information enhanced schema-linking substantially improve the performance of our model. At the time of writing, our model ranks first on both two large-scale cross-domain context-dependent text-to-SQL leaderboards, \sparc{}~\citep{DBLP:conf/acl/YuZYTLLELPCJDPS19} and \cosql{}~\citep{DBLP:conf/emnlp/YuZELXPLTSLJYSC19}. Specifically, our model achieves a $64.6\%$ question match and $42.9\%$ interaction match accuracy on \sparc{}, and a $53.9\%$ question match and $24.6\%$ interaction match accuracy on \cosql{}.

\section{Related Work}

Text-to-SQL semantic parsing follows a long line of research on semantic parsing from natural language to logical language~\citep{DBLP:conf/aaai/ZelleM96,DBLP:conf/uai/ZettlemoyerC05,DBLP:conf/acl/WongM07}.

Recently, \cits{} semantic parsing has been well studied.~\spider{}~\citep{DBLP:conf/emnlp/YuZYYWLMLYRZR18}
is a famous dataset for the complex and cross-domain \cits{} task.
Some works~\citep{DBLP:conf/acl/BoginBG19,DBLP:conf/emnlp/BoginGB19,DBLP:conf/naacl/ChenCZCXZY21} apply graph neural networks to encode database schema.~\citet{DBLP:conf/acl/Xu0YZT0CPC20} succeed in appling deep transformers to the \cits{} task.~\citet{DBLP:conf/emnlp/YuYYZWLR18} employ a tree-based decoder to match SQL grammar.~\citet{DBLP:conf/naacl/RubinB21} improve the tree-based decoder by a bottom-up method.~\citet{scholak2021picard} refine the sequence-based decoder via carefully designed restriction rules.~\citet{DBLP:conf/acl/GuoZGXLLZ19} and~\citet{DBLP:journals/corr/abs-2109-05153} propose SQL intermediate representations to bridge the gap between natural language and SQL.~\citet{DBLP:conf/emnlp/LeiWMGLKC20} study the role of schema-linking in text-to-SQL semantic parsing.~\citet{DBLP:conf/acl/WangSLPR20} propose a unified framework to capture the schema-linking.~\citet{DBLP:conf/emnlp/LinSX20} represent the schema-linking as a tagged sequence.~\citet{DBLP:conf/acl/CaoC0ZZ020} further integrate non-local and local features via taking advantage of both schema-linking graph and its corresponding line graph. Besides, many previous works~\citep{DBLP:conf/naacl/DengAMPSR21,DBLP:conf/iclr/0009WLWTYRSX21,DBLP:conf/aaai/ShiNWZLWSX21} focus on pre-train models for \cits{} semantic parsing.

With more attentions on \cdts{} semantic parsing, existing works have been devoted to the \cdts{} task. \sparc~\citep{DBLP:conf/acl/YuZYTLLELPCJDPS19} and \cosql~\citep{DBLP:conf/emnlp/YuZELXPLTSLJYSC19} datasets are specially proposed for the task. \editsql~\citep{DBLP:conf/emnlp/ZhangYESXLSXSR19} and \istsql~\citep{DBLP:conf/aaai/WangLZ021} focus on taking advantages of the last predicted query for the prediction of current query.
\editsql{} tries to copy the overlap tokens from the last predicted query, while \istsql{} proposes an interaction state tracking method to encode the information from the last predicted query.
\igsql~\citep{DBLP:conf/emnlp/CaiW20} and \rtsql~\citep{DBLP:conf/aaai/HuiGRLLSHSZZ21} leverages the contextual information among the current utterance, interaction history utterances and database schema via context-aware dynamic graphs. Notably, \rtsql{} simulates the information by connecting the schema graphs with the tokens in interactive utterances.~\citet{DBLP:conf/iclr/0009ZPMA21} creatively propose a context-aware pre-trained language model. However, the problem of making full use of both interaction history utterances and predicted queries for the \cdts{} task remains open.

\section{HIE-SQL}
First, we formally define the conversational text-to-SQL semantic parsing problem. In the rest of the section, we detail the architecture of history information enhanced text-to-SQL model (HIE-SQL).

\subsection{Preliminaries}
\paragraph{\textbf{Task Definition.}} Given the current user utterance $u_\tau$, interaction history $h_\tau = [u_1, u_2, ..., u_{\tau-1}]$, the schema $D = \langle T,C\rangle$ of the target database such that the set of tables $T =\{t_1,...,t_{|T|}\}$ and the set of columns $C = \{c_1,...,c_{|C|}\}$, our goal is to generate the corresponding SQL query $s_\tau$.

\begin{figure}[t]
    \centering
    \includegraphics[width=0.49\textwidth]{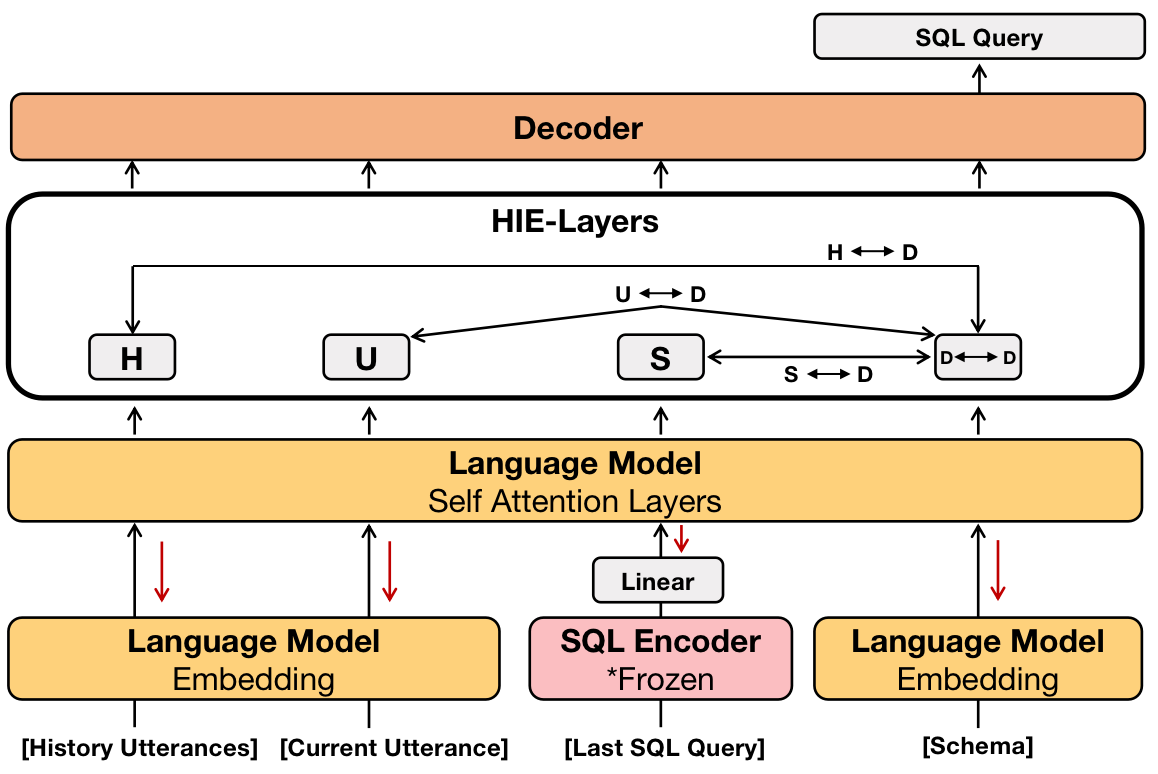}
    \caption{Structure and components of HIE-SQL. The red arrows represent the direction of back propagation during the training stage, witch means parameters of SQL Encoder will not be updated during training. Linear represents one fully connected layer. And we use SQLBERT as the SQL Encoder in the structure.}
    \label{fg:popg2}
\end{figure}

\paragraph{\textbf{Model Architecture.}} Figure~\ref{fg:popg2} shows the encoder-decoder framework of HIE-SQL. We will introduce it in four modules:
\begin{enumerate*}[label=(\roman*)]
    \item \textbf{Multimodal Encoder}, which encodes SQL query and natural language context in a multimodal manner,
    \item \textbf{SQLBERT}, a bimodal pre-trained encoder for SQL and natural language,
    \item \textbf{HIE-Layers}, which encode pre-defined schema-linking relations between all elements of the output of Language Model, and
    \item \textbf{Decoder}, which generates SQL query as an abstract syntax tree.
\end{enumerate*}

\subsection{Multimodal Encoder}
Since of the huge syntax structure differences between SQL and natural language, using a single language model to encode both languages at the same time increases the difficulty and cost of training the model. Inspired by the efficiency of the works~\cite{DBLP:conf/nips/KielaBFT19,DBLP:journals/corr/abs-2106-13884} to solve the multimodal problems, we build an additional pre-trained Encoder named SQLBERT (we will detail it in the following section) to pre-encode SQL query. 
Then we learn weights $W \in R^{N\times M}$ to project the N-dimensional SQL query embeddings to M-dimensional token input embedding space of the language model: 
\begin{equation}\label{e1}
\mathcal{S} = Wf(s_{\tau-1}),
\end{equation}
where $f(\cdot)$ is the last hidden state output of SQLBERT.

We arrange the input format of HIE-SQL as $x= (\texttt{\textup{[CLS]}},\mathcal{U},\texttt{\textup{[CLS]}},\mathcal{S},\texttt{\textup{[SEP]}},\mathcal{T},\texttt{\textup{[SEP]}},\mathcal{C})$ in which
\begin{equation}
\begin{aligned}
\mathcal{U} &= (u_1,\texttt{\textup{[CLS]}},u_2,...,\texttt{\textup{[CLS]}},u_\tau), \\
\mathcal{T} &= (t_1,\texttt{\textup{[SEP]}},t_2,...,\texttt{\textup{[SEP]}},t_{|T|}), \\
\mathcal{C} &= (c_1,\texttt{\textup{[SEP]}},c_2,...,\texttt{\textup{[SEP]}},c_{|C|}).
\end{aligned}
\end{equation}
All the special separator tokens and language word tokens in $x$ are converted to the word embedding by embedding layer of the language model. Gathering the embeddings of natural language and SQL, we feed them to self-attention blocks in a language model. In the training stage, we directly take the golden SQL query of the last turn as an input SQL query and set $\mathcal{S}$ to empty for the first turn. As for the inference stage, we apply the SQL query generated by HIE-SQL in the last turn.

\subsection{SQLBERT}\label{SQLBERT}

As mentioned above, we treat the SQL query as another modality that can provide information of the SQL query from the previous round as a reference for the model. So we need an encoder to extract the representation of the SQL query.

\paragraph{\textbf{Model Architecture.}} Considering the success of multi-modal pre-trained models, such as ViLBERT~\citep{DBLP:conf/nips/LuBPL19} for language-image and CodeBERT~\citep{DBLP:conf/emnlp/FengGTDFGS0LJZ20} for natural language and programming language, we propose SQLBERT, a bimodal pre-trained model for natural language and SQL. We develop SQLBERT by using the same model architecture as RoBERTa~\citep{liu2019roberta}. The total number of model parameters is 125M.

\begin{table*}
\newcommand{\tabincell}[2]{\begin{tabular}{@{}#1@{}}#2\end{tabular}}
\centering
\setlength{\tabcolsep}{7mm}{
\begin{tabular}{c|c|c|c}
\hline
& Current Utterance & Interaction History & SQL Query \\
\hline
 Columns & \tabincell{l}{U$-$C$-$EM \\ U$-$C$-$PM \\ U$-$C$-$VM} & \tabincell{l}{H$-$C$-$EM \\ H$-$C$-$PM \\ H$-$C$-$VM} & \tabincell{l}{S$-$C$-$EC\\ S$-$C$-$UC} \\
\hline
 Tables & \tabincell{l}{U$-$T$-$EM \\ U$-$T$-$PM} & \tabincell{l}{H$-$T$-$EM \\ H$-$T$-$PM} & \tabincell{l}{S$-$T$-$ET \\ S$-$T$-$UT} \\
 \hline
\end{tabular}}

\caption{Edge types between current utterance $U$, interaction history $H$, SQL $S$, and database schema $D$ (Columns $C$ and Tables $T$). We set two match types between the language tokens of $U$, $H$, and $D$: \textbf{EM} for Exact Match, \textbf{PM} for Partial Match. When using database contents, we set \textbf{VM} (Value Match) for exactly matching the value of columns. As for SQL $S$, we simply match the words of tables and columns that appear in it to the target database schema: \textbf{EC} (Equal Columns) and \textbf{UC} (Unequal Columns) for columns, \textbf{ET} (Equal Tables) and \textbf{UT} (Unequal Tables) for tables. And we omit the pre-existing relations in schema such as the foreign-key relation (C-C-FK) in the table.}
\label{graphlink}
\end{table*}

\begin{figure}[t]
    \centering
    \includegraphics[width=0.46\textwidth]{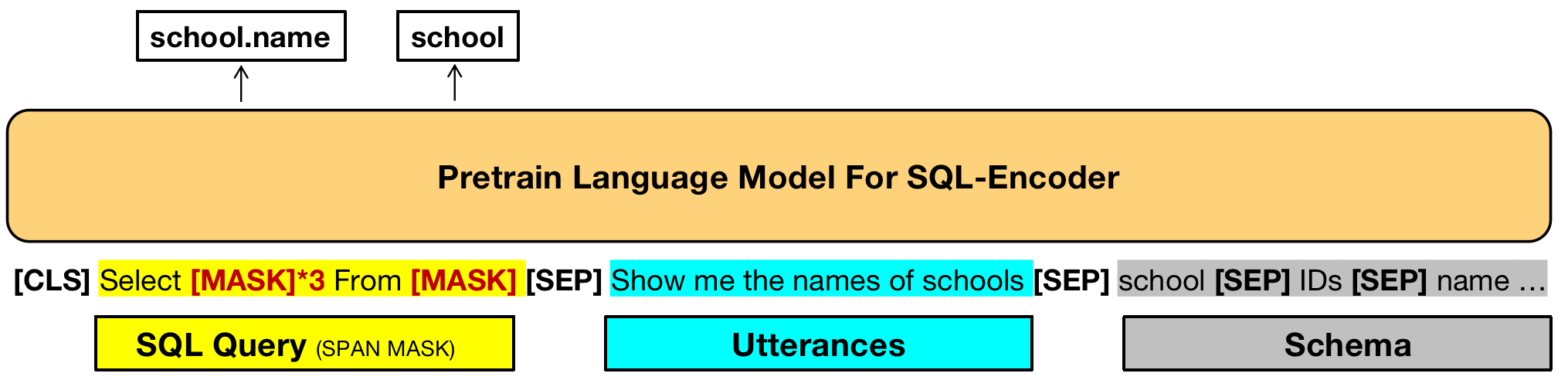}
    \caption{Input format and training objective of SQLBERT.}
    \label{fg:popg3}
\end{figure}

\paragraph{\textbf{Input Format.}} As the training method showed in Figure~\ref{fg:popg3}, we set the same input as CodeBERT~\citep{DBLP:conf/emnlp/FengGTDFGS0LJZ20} does. To alleviate the difficulty of training and resolve inconsistencies between natural language and schema, we append the question-relevant database schema to the concatenation of SQL query and question. We represent the whole input sequence into the format as $x =$ $(\texttt{\textup{[CLS]}},s_1,s_2,..s_n,\texttt{\textup{[SEP]}},q_1,q_2,..q_m,\texttt{\textup{[SEP]}},$ $t_1:c_{11},c_{12},...,\texttt{\textup{[SEP]}},t_2:c_{21},...,\texttt{\textup{[SEP]}},...)$, in which $s$, $q$, $t$, and $c$ are the tokens of SQL query, question, tables, and columns respectively. 

\paragraph{\textbf{Training Objective.}} The main training objective of SQLBERT is the masked language modeling (MLM). It's worth noting that we only mask the tokens of SQL query because we only need SQLBERT to encode SQL query in the downstream task. Specifically, we utilize a special objective referenced span masking~\citep{sun2019ernie} by sampling 15\% independent span in SQL clause except the reserved word (e.g., SELECT, FROM, WHERE), which aims to avoid leaking answers and help SQLBERT learn the information structure of SQL better. In the training stage, we adopt a dynamic masking strategy via randomly shuffling the order of tables and columns in the original schema. We describe the masked span prediction loss as
\begin{equation}\label{e0}
\mathcal{L}(\theta) = \sum_{k=1}^n-log\mathcal{P}_\theta(s_k^{mask}|s^{\backslash mask},q,t,c),
\end{equation}
where $\theta$ stands for the model parameters, $s_k^{mask}$ is the masked span of SQL input, $s^{\backslash mask}$ is the unmasked part.

\begin{figure}[t]
    \centering
    \includegraphics[width=0.47\textwidth]{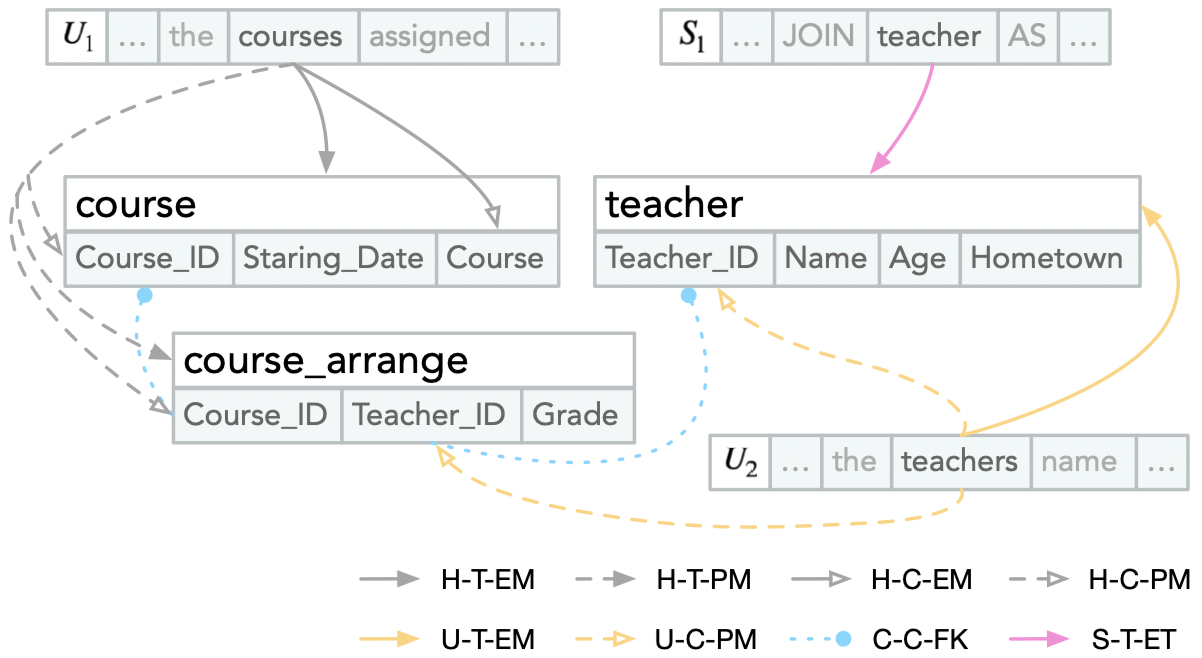}
    \caption{An example of the schema-linking graph for the prediction of $S_2$ in Figure~\ref{fg:popg1}. The graph is a subgraph of the whole schema-linking graph. We only respectively choose one token in the history utterance ($U_1$), the current utterance ($U_2$), and the last predicted SQL query ($S_1$) in the example. Besides, we omit all unequal relation edges (S-C-UC and S-T-UT) and default "no relation" edges.}
    \label{fg:schema-linking}
\end{figure}

\paragraph{\textbf{Training Data.}} Unlike \score{}~\citep{DBLP:conf/iclr/0009ZPMA21}, which uses multiple open-source text-to-SQL datasets (W{\scriptsize{IKI}}T{\scriptsize{ABLES}}~\citep{DBLP:conf/semweb/BhagavatulaND15}, \wikisql{}~\citep{DBLP:journals/corr/abs-1709-00103}, \spider{}, \sparc{}, and \cosql{}) and data synthesis methods to obtain a large amount of pre-training data, we train SQLBERT only with the datasets including \spider{}, \sparc{} and \cosql{}. For each sample, we only use its question, SQL query, and the corresponding database schema. As for \sparc{} and \cosql{}, which is a context-dependent version, we simply concatenate the current utterance with the history utterances to build the question input. The size of the training dataset is 34,175.

\subsection{HIE-Layers}\label{hie-layer}
\paragraph{\textbf{Schema-Linking Graph.}} To explicitly encode the complex relational database schema. We convert it to a directed graph $\mathcal{G} = \langle\mathcal{V},\mathcal{E}\rangle $, where $\mathcal{V} = C \cup T $ and $\mathcal{E}$ represents the set of pre-existing relations within columns and tables such as the foreign-key relation.
In addition, we also consider the unseen linking to the schema in the contexts of current utterance, interaction history utterances, and the last predicted SQL query. Specifically, we define the context-dependent schema-linking graph $\mathcal{G}_c = \langle\mathcal{V}_c,\mathcal{E}_c\rangle$ where $\mathcal{V}_c = C \cup  T \cup  U \cup  H \cup  S$ and $\mathcal{E}_c = \mathcal{E} \cup  \mathcal{E}_{U\leftrightarrow D} \cup  \mathcal{E}_{H\leftrightarrow D} \cup  \mathcal{E}_{S\leftrightarrow D}$. The additional relation edges are listed in Table~\ref{graphlink}. In Figure~\ref{fg:schema-linking}, we show an example of the proposed schema-linking graph. 

\paragraph{\textbf{Graph Encoding.}} The work~\citep{DBLP:conf/acl/WangSLPR20} shows that Relative Self-Attention Mechanism~\citep{DBLP:conf/naacl/ShawUV18} is an efficient way to encode graphs whose nodes are at the token level. It rebuilds the calculation of the self-attention module in the transformer layers as follows:

\begin{equation}
\begin{aligned}
e_{ij}&=\frac{x_iW^Q(x_jW^K +  \textcolor{red}{r_{ij}^K})^T}{\sqrt{d_z}}, \\
\alpha_{ij} &= \mathop{softmax}\limits_{j}\{e_{ij}\}, \\
z_i &= \sum_{j=1}^n\alpha_{ij}(x_jW^V + \textcolor{red}{r_{ij}^V}).
\end{aligned}
\label{e3}
\end{equation}

HIE-Layers consist of 8 transformer layers, whose self-attention modules are described above. Specifically, we initialize a learned embedding for each type of edge defined above. For every input sample, we build a relation matrix $\mathcal{R} \subseteq (L \times L)$ where L is the length of the input token. $\mathcal{R}^{(i,j)}$ represents the relation type between $i-$th and $j-$th input tokens. While computing the relative attention, we set the $r_{ij}^K = r_{ij}^V = \mathcal{R}_e^{(i,j)} $ where $\mathcal{R}_e^{(i,j)}$ is the corresponding embedding of $\mathcal{R}^{(i,j)}$.

\begin{table*}
\centering
\newcommand{\tabincell}[2]{\begin{tabular}{@{}#1@{}}#2\end{tabular}}
\begin{tabular}{ccccccccccccc}
\toprule
Dataset & \tabincell{c}{System \\ Response} & Interaction & Train & Dev & Test & User Questions & Vocab & Avg Turn \\
\midrule
 \cosql{} & \CheckmarkBold & 3007 & 2164 & 293 & 551 & 15598 & 9585 & 5.2 \\
 \sparc{} & \XSolidBrush & 4298 & 3034 & 422 & 842 & 12726 & 3794 & 3.0  \\
 \bottomrule
\end{tabular}
\caption{Details of \sparc{} and \cosql{} datasets.}
\label{dataset}
\end{table*}

\begin{table*}
    \centering
    \begin{tabular}{lcccccccc}
    \toprule
         & \multicolumn{4}{c}{\textbf{\sparc{}}} & \multicolumn{4}{c}{\textbf{\cosql{}}} \\
         \cmidrule(lr){2-5}\cmidrule(lr){6-9}
         \multicolumn{1}{c}{\textbf{Model}} & \multicolumn{2}{c}{\textbf{Dev}} & \multicolumn{2}{c}{\textbf{Test}} & \multicolumn{2}{c}{\textbf{Dev}} & \multicolumn{2}{c}{\textbf{Test}} \\ 
         \cmidrule(lr){2-3}\cmidrule(lr){4-5}\cmidrule(lr){6-7}\cmidrule(lr){8-9}
         & \textbf{QM} & \textbf{IM} & \textbf{QM} & \textbf{IM} & \textbf{QM} & \textbf{IM} & \textbf{QM} & \textbf{IM} \\
    \toprule
    \editsql{} + BERT~\citep{DBLP:conf/emnlp/ZhangYESXLSXSR19} & 47.2 & 29.5 & 47.9 & 25.3 & 39.9 & 12.3 & 40.8 & 13.7 \\
    \igsql{} + BERT~\citep{DBLP:conf/emnlp/CaiW20} & 50.7 & 32.5 & 51.2 & 29.5 & 44.1 & 15.8 & 42.5 & 15.0 \\
    \istsql{} + BERT~\citep{DBLP:conf/aaai/WangLZ021} & - & - & - & - & 44.4 & 14.7 & 41.8 & 15.2 \\
    \rtsql{} + BERT~\citep{DBLP:conf/aaai/HuiGRLLSHSZZ21} & 54.1 & 35.2 & 55.8 & 30.8 & 45.7 & 19.5 & 46.8 & 17.0 \\
    RAT-SQL$^\ttd$ + \score{}~\citep{DBLP:conf/iclr/0009ZPMA21} & 62.2 & 42.5 & 62.4 & 38.1 & 52.1 & 22.0 & 51.6 & 21.2 \\
    T5-3B + \picard{}$^\ttd$~\citep{scholak2021picard} & - & - & - & - & \textbf{56.9} & 24.2 & \textbf{54.6} & 23.7 \\
    \midrule
    HIE-SQL + \grappa{} (ours) & \textbf{64.7} & \textbf{45.0} & \textbf{64.6} & \textbf{42.9} & 56.4 & \textbf{28.7} & 53.9 & \textbf{24.6} \\
    \bottomrule
    \end{tabular}
    \caption{Performances of various models in \sparc{} and \cosql{}. QM and IM stand for question match and interaction match respectively. The models with $\ttd$ are proposed for the \cits{} task and applied to the \cdts{} task by just appending interaction history utterances to the input.}
    \label{result}
\end{table*}

\subsection{Decoder}
To build the decoder of HIE-SQL, we apply the same work~\citep{DBLP:conf/acl/YinN17} as~\citet{DBLP:conf/acl/WangSLPR20} propose, which generates SQL as an abstract syntax tree in depth-first traversal order by using LSTM~\citep{DBLP:journals/neco/HochreiterS97} to output sequences of decoder actions. We recommend the reader to refer to the work~\citep{DBLP:conf/acl/YinN17} for details.

\subsection{Regularization Strategy}
We introduce R-Drop~\citep{DBLP:journals/corr/abs-2106-14448}, a simple regularization strategy, to prevent the overfitting of the model.
Concretely, we feed every input data $x_i$ to go through our model twice and the loss function is as follows:
\begin{equation}
\begin{aligned}
\mathcal{L}_{NLL}^i =& -log\mathcal{P}_1(y_i|x_i) - log\mathcal{P}_2(y_i|x_i), \\
\mathcal{L}_{KL}^i =&~\frac{1}{2}(D_{KL}(\mathcal{P}_1(y_i|x_i)\| \mathcal{P}_2(y_i|x_i)) \\ 
&+ D_{KL}(\mathcal{P}_2(y_i|x_i)\| \mathcal{P}_1(y_i|x_i))), \\
\mathcal{L}^i =&~\mathcal{L}_{NLL}^i + \mathcal{L}_{KL}^i,
\end{aligned}
\label{e4}
\end{equation}
where -$log\mathcal{P}_1(y_i|x_i)$ and -$log\mathcal{P}_2(y_i|x_i)$ are two output distributions for input $x_i$ at all decoder steps, $\mathcal{L}_{NLL}^i$ is the negative log-likelihood learning objective of decoder actions, and $\mathcal{L}_{KL}^i$ is the bidirectional Kullback-Leibler (KL) divergence between these two output distributions.

\section{Experiment}
\subsection{Setup}
\noindent{\textbf{Setting.}} We initialize the weights of Language Model with \grappa{}~\citep{DBLP:conf/iclr/0009WLWTYRSX21}, an effective pre-training model for table semantic parsing that performs well on the context-independent text-to-SQL datasets (e.g. Spider). We
stack 8 HIE-layers, which are introduced in section~\ref{hie-layer}, on top of the Language Model. When training the model with R-Drop, we set the Dropout rate of 0.1 for the Language Model and HIE-Layers, 0.3 for the decoder.
We use Adam optimizer to conduct the parameter learning and set the learning rate of $1e^{-5}$ for fine-tuning \grappa{} and $1e^{-4}$ for HIE-Layers and Decoder. The learning rate  linearly increases to the setting point at first $max\_steps / 8$ steps, then decreases to 0 at $max\_steps$, where $max\_steps = 50000$ with 24 training batch-size. As for SQLBERT, we fine-tune CodeBERT$_{BASE}$~\citep{DBLP:conf/emnlp/FengGTDFGS0LJZ20} on the dataset we described in Section~\ref{SQLBERT}. We set the learning rate as $1e^{-5}$, a batch size of 64, and train SQLBERT for 10 epochs. The shape of learned weights of the linear layer applied to the output of SQLBERT is $768\times1024$. We only need one V100 (32G) GPU to train our model. While inferring, we set the beam size to 3.

\noindent{\textbf{Datasets.}} We conduct experiments on two cross-domain context-dependent text-to-SQL datasets, \sparc{}~\citep{DBLP:conf/acl/YuZYTLLELPCJDPS19} and \cosql{}~\citep{DBLP:conf/emnlp/YuZELXPLTSLJYSC19}. Table~\ref{dataset} depicts the statistic information of them.

\begin{figure*}[t]
    \centering
    \includegraphics[width=0.9\textwidth]{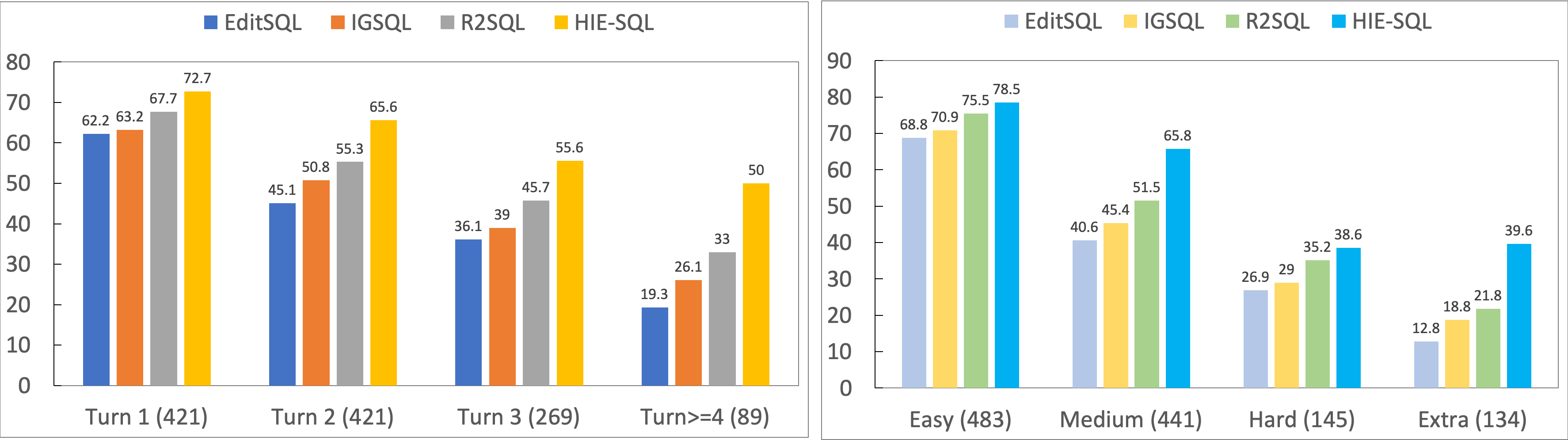}
    \caption{Performances of previous works and HIE-SQL in different turns (left) and different difficulty levels (right) on \sparc{}.}
    \label{fg:popg4}
\end{figure*}

\noindent{\textbf{Evaluation Metrics.}} The main metric we used to measure model performance in \sparc{} and \cosql{} is interaction match (IM), which requires all output SQL queries in interaction to be correct. We also use question match (QM) to evaluate the accuracy of every single question. 

\subsection{Experiment Result}

Results of our proposed HIE-SQL model are shown in Table~\ref{result}. In terms of interaction match, our model achieves state-of-the-art performances on both development set and test set of \sparc{} and \cosql{}. For the test set of \sparc{}, HIE-SQL outperforms the prior state-of-the-art~\citep{DBLP:conf/iclr/0009ZPMA21} by 4.8$\%$ in IM and 2.2$\%$ in QM. For \cosql{}, compared with the previous state-of-the-art~\citep{scholak2021picard}, a rule-based auto-regressive method based on the large pre-trained model-T5-3B~\citep{DBLP:journals/jmlr/RaffelSRLNMZLL20} which contains 2.8 billion parameters, HIE-SQL improves IM of development set by 4.5$\%$ and IM of the test set by 0.9$\%$ with only 580M parameters. Besides, HIE-SQL surpasses RAT-SQL + \score{} in all metrics of \sparc{} and \cosql{}. This demonstrates that properly integrating interaction utterances and predicted SQL queries is an effective way to enhance the model's ability for Context-Dependent Text-to-SQL Semantic Parsing. 

To further explore the advantages of HIE-SQL, we test the performance on different turns and at different difficulty levels of utterances. As shown in Figure~\ref{fg:popg4}, with the increase of turns, the lead of our model gets greater and greater. When the indexes of turns are greater than or equal to 4, the accuracy of HIE-SQL is 17\% higher than that of \rtsql{}. It demonstrates that the main contribution of introducing SQL query is to improve the robustness of the model to long 
interaction.
HIE-SQL is also robust to the varying difficulty levels of utterances. Our model performs equally in hard and extra hard levels, and achieves 39.6\% accuracy on the extra hard level, which is 17.8\% higher than that of \rtsql{}.

\begin{table}[t]
    \centering
    \begin{tabular}{lcccc}
    \toprule
         & \multicolumn{2}{c}{\textbf{\sparc{}}} & \multicolumn{2}{c}{\textbf{\cosql{}}} \\
    \cmidrule(lr){2-3}\cmidrule(lr){4-5}
    \textbf{Model} & \textbf{QM} & \textbf{IM} & \textbf{QM} & \textbf{IM} \\
    \toprule
    HIE-SQL & 64.7 & \textbf{45.0} & 56.4 & \textbf{28.7} \\
    \midrule
    \setlength{\parindent}{4em} w/o SQL query & \textbf{65.8} & 44.3 & \textbf{56.5} & 23.9 \\
    \setlength{\parindent}{4em} w/o SQLBERT & 63.9 & 44.7 & 54.8 & 26.3 \\
    \setlength{\parindent}{4em} w/o $\mathcal{E}_{H\leftrightarrow D}$ & 64.0 & 44.3 & 56.0 & 26.3 \\
    \bottomrule
    \end{tabular}
    \caption{Ablation study of HIE-SQL in development sets of \sparc{} and \cosql{}. As for ablation on SQL query, we drop the SQL query and only feed utterances and database schema to the model. As for ablation on SQLBERT, we directly concatenate the tokens of SQL query and other context tokens for the input of the language model. And w/o $\mathcal{E}_{H\leftrightarrow D}$ means we treat historical utterances like the current utterance in our schema-linking.}
    \label{ablation}
\end{table}


\begin{table}[t]
    \centering
    \begin{tabular}{c|c|c|c|c}
    \hline
    Dataset & Model & T-F & F-T & T-T \\
    \hline
    \multirow{2}*{\sparc{}} & HIE-SQL & 125 & 88 & \textbf{383} \\
    \cline{2-5}
    & w/o SQL query & 132 & 104 & 379 \\
    \hline
    \multirow{2}*{\cosql{}} & HIE-SQL & 140 & 106 & \textbf{278} \\
    \cline{2-5}
    & w/o SQL query & 161 & 128 & 254 \\
    \hline
    \end{tabular}
    \caption{The counts of different switches in the pairs of adjacent predicted SQL queries. T-F stands for the match of the former predicted query and unmatch of the later predicted query with golden queries. F-T stands for the reverse case. T-T is the case of both matching.}
    \label{adjacent}
\end{table}

\subsection{Ablation Study}
We provide ablation studies to examine the contribution of each component of HIE-SQL. We want to identify whether introducing the last SQL query has a significant impact on performance. Also, we would like to investigate whether the pre-trained SQL encoder, SQLBERT, can improve the model's ability to understand SQL queries. What's more, we conduct another ablation study regarding additional graph edges between historical utterances and database schema $\mathcal{E}_{H\leftrightarrow D}$ to check the necessity of the join of historical utterance information in schema-linking.

As shown in Table~\ref{ablation}, Our full model achieves about 5 points and 1 point improvement of IM in \cosql{} and \sparc{} respectively compared with the model without the last SQL query input. The pre-encoding SQL query by SQLBERT can further improve the performance. It confirms SQLBERT's ability to efficiently represent SQL features. In addition, $\mathcal{E}_{H\leftrightarrow D}$ also plays a positive role.

Table~\ref{adjacent} shows the continuity of performance of our model compared with that of the model without the last SQL query input. Our model has a higher rate of continuous match, but a lower rate of switching from mismatch to match. It illustrates that our model does use the SQL information and is sensitive to the accuracy of the last predicted SQL query which explains the higher question match without SQL query input.

\begin{figure}[t]
    \centering
    \includegraphics[width=0.45\textwidth]{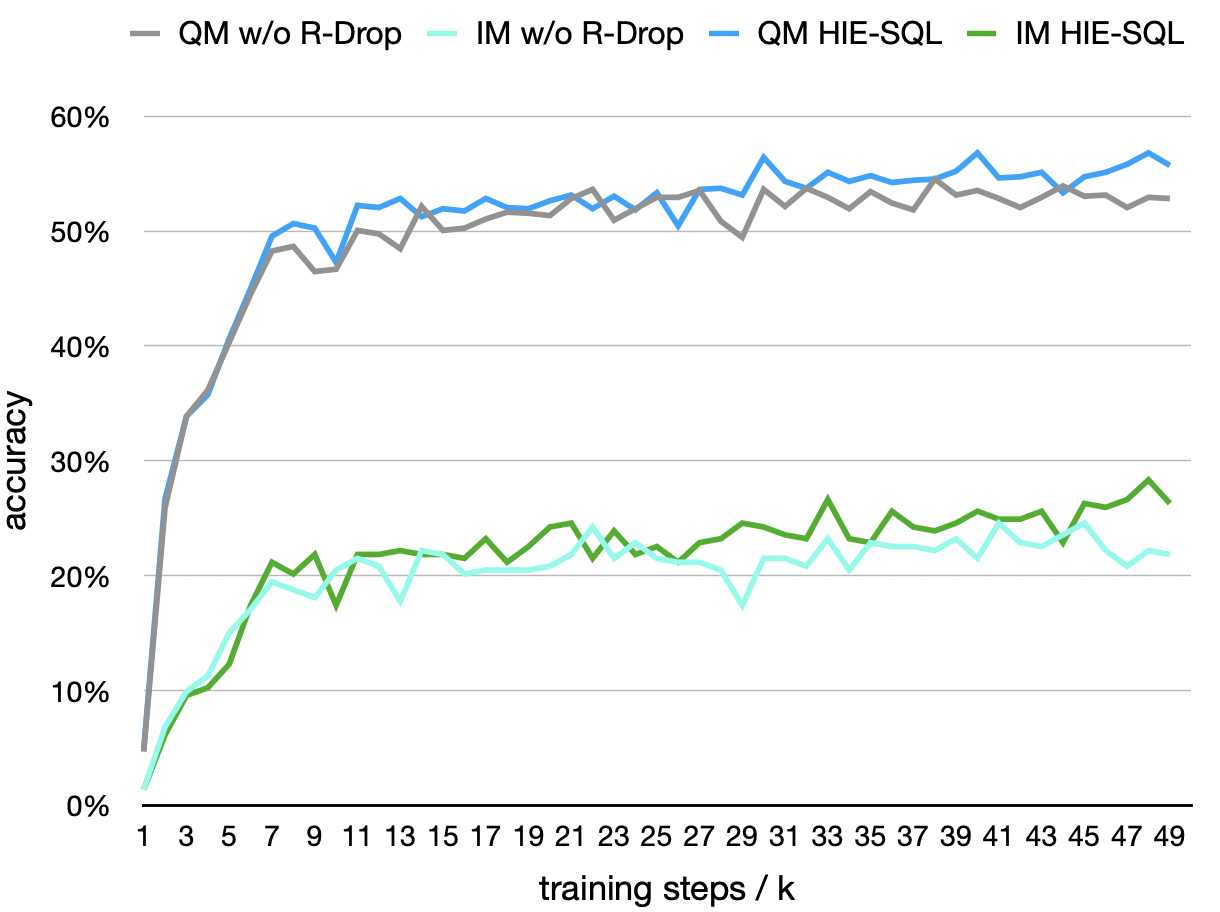}
    \caption{Ablation study result of regarding R-Drop in development set of \cosql{}. We show the performances in QM and IM of two models at different training steps. We set the beam size = 1 in the inference stage.}
    \label{R-Drop}
\end{figure}

\begin{table*}[t]
\centering
\newcommand{\tabincell}[2]{\begin{tabular}{@{}#1@{}}#2\end{tabular}}

\begin{tabular}{rl}

\toprule
$U_{a1}$ & Which cartoon aired {\color{red}first}? \\
\midrule
HIE-SQL & SELECT title FROM cartoon ORDER BY original\_air\_date asc LIMIT 1 \\
RAT-SQL & SELECT title FROM cartoon ORDER BY original\_air\_date asc LIMIT 1 \\
\midrule
$U_{a2}$ & What was the {\color{red}last} cartoon to air? \\
\midrule
HIE-SQL & SELECT title FROM cartoon ORDER BY original\_air\_date desc LIMIT 1 \\
RAT-SQL & SELECT title FROM cartoon ORDER BY original\_air\_date desc LIMIT 1 \\
\midrule
$U_{a3}$ & What channel was it on? \\
\midrule
HIE-SQL & SELECT channel FROM cartoon ORDER BY original\_air\_date desc LIMIT 1 \\
RAT-SQL & SELECT channel FROM cartoon ORDER BY original\_air\_date desc LIMIT 1 \\
\midrule
$U_{a4}$ & What is the production code? \\
\midrule
HIE-SQL & SELECT production\_code FROM cartoon ORDER BY original\_air\_date {\color{red}desc} LIMIT 1 \\ 
RAT-SQL & SELECT production\_code FROM cartoon ORDER BY original\_air\_date {\color{blue}asc} LIMIT 1 \\
\bottomrule

~\\

\toprule
$U_{b1}$ & List the name of the teachers and the courses assigned for them to teach. \\
\midrule
HIE-SQL & SELECT Name, Course FROM ... \\
RAT-SQL & SELECT Name, Course FROM ... \\
\midrule
$U_{b2}$ & Arrange this list with the {\color{red}teachers name} in ascending order \\
\midrule
HIE-SQL & ELECT Name, Course FROM ... ORDER BY {\color{red}Name} Asc \\
RAT-SQL & ELECT Name, Course FROM ... ORDER BY {\color{red}Name} Asc \\
\midrule
$U_{b3}$ & {\color{red}Include teachers ID} in tha same list \\
\midrule
HIE-SQL & SELECT {\color{red}Name, Course, Teacher\_ID} FROM ... ORDER BY {\color{red}Name} Asc \\
RAT-SQL & SELECT {\color{blue}Teacher\_ID} FROM ... ORDER BY {\color{blue}Teacher\_ID} Asc \\
\bottomrule

\end{tabular}

\caption{Examples in \cosql{}. $U_{ij}$ is the input utterance of turn $j$ of example $i$ with corresponding predictions of HIE-SQL and RAT-SQL following. All predictions of HIE-SQL are the ground truth queries in the cases.}
\label{case_study}
\end{table*}

As shown in Figure~\ref{R-Drop}, the model with R-drop outperforms the model without R-Drop in both QM and IM. Additionally, the standard deviations of the IM in the last 20k steps are 0.014 and 0.015 of HIE-SQL and the one without R-Drop respectively even the curve of HIE-SQL has a more obvious upward trend. It shows that R-Drop improves the robustness of our model and stabilizes its performance in IM. What's more, when the key information the last SQL query is introduced, our model needs more training steps to fit the same training data. After adding R-drop, in the same training step, the model will forward the data twice to get the KL loss. This is equivalent to doubling the amount of training data in the same step. Therefore, our model has learned the training data more fully and is able to make full use of various historical interaction information.

\subsection{Case Study}

In Table~\ref{case_study}, we offer some cast studies about the performance of HIE-SQL and RAT-SQL in the examples of \cosql{} in order to demonstrate the superiority of HIE-SQL in \cdts{} semantic parsing problems more visually. As the examples show, RAT-SQL fails to distinguish the right one from two long-range dependences in $U_{a1}$ and $U_{a2}$ in the first example and fails to inherit the query information from $U_{b2}$ in $U_{b3}$. By contrast, HIE-SQL inherits the right context-dependence from the last predicted query to avoid the confusion.

\section{Conclusion}
We present HIE-SQL, a history information enhanced \cdts{} model, which targets at explicitly capturing the context-dependence from both interaction history utterances and the last predicted SQL query. With the help of the proposed bimodal pre-trained model, SQLBERT, HIE-SQL bridge the gap between the utterances and predicted SQL despite the mismatch of natural language and logic-form SQL. Moreover, we also introduce a method of schema-linking to enhance the connections among utterances, SQL query, and database schema.

Taken together, HIE-SQL achieves consistent improvements on the \cdts{} task, especially in the interaction match metric. HIE-SQL achieves new state-of-the-art results on two famous \cdts{} datasets, \sparc{} and \cosql{}.

\small
\bibliography{anthology,acl2020}
\bibliographystyle{acl_natbib}

\end{document}